\documentclass[11pt,a4paper]{article}
\usepackage{jheppub}
\usepackage[utf8]{inputenc}
\usepackage{amsmath}
\usepackage{amsthm}
\usepackage{amssymb}
\usepackage{enumitem}   
\usepackage[english]{babel}
\usepackage{url}
\usepackage{mathtools}
\usepackage{bbold}
\usepackage{slashed}

\usepackage{subcaption}
\usepackage{tikz}
\usetikzlibrary{arrows,shapes.misc,decorations.markings,decorations.pathmorphing,positioning,intersections}
\tikzset{cross/.style={cross out, draw=black, minimum size=2*(#1-\pgflinewidth), inner sep=0pt, outer sep=0pt},
cross/.default={1.5mm}}
\tikzset{mydash/.style={dashed, dash pattern=on 4pt off 5pt}}
\tikzset{
  vertex/.style={draw,shape=circle,fill=black,minimum size=3pt,inner sep=0pt},
  cross/.style={cross out, draw=black,thick, minimum size=6pt, inner sep=0pt, outer sep=0pt},
  external/.style={inner sep=2pt},
  plabel/.style={inner sep=2pt},
  blob/.style={circle,fill=black!20,minimum size=0.7cm,draw,thick},
  whiteblob/.style={circle,fill=white,minimum size=1.0cm,draw,thick},
  effective/.style={rectangle,fill=black!20,minimum size=0.5cm,draw,thick},
  vev/.style={shape=vev,draw,inner sep=2pt,thick},
  mass/.style={shape=cross,draw,thick},
  rscalar/.style={dashed,thick},
  mfermion/.style={thick},
  scalar/.style={postaction={decorate}, decoration={markings,mark=at position .55 with {\arrow{latex}}},dashed,thick},
  ooscalar/.style={postaction={decorate}, decoration={markings,mark=at position .7 with {\arrow{latex}}},dashed,thick},
  fermion/.style={postaction={decorate}, decoration={markings,mark=at position .55 with {\arrow{latex}}},thick},
  majfermion/.style={postaction={decorate}, decoration={markings,mark=at position .7 with {\arrow{latex}}},thick},
  oofermion/.style={postaction={decorate}, decoration={markings,mark=at position .85 with {\arrow{latex}}, mark=at position .35 with {\arrowreversed{latex}}},thick},
  iifermion/.style={postaction={decorate}, decoration={markings,mark=at position .35 with {\arrowreversed{latex}}, mark=at position .85 with {\arrow{latex}}},thick},
  gaugeboson/.style={decorate, decoration={snake},thick},
  gluon/.style={decorate, decoration={coil,amplitude=4pt, segment length=5pt},thick},
  photon/.style={decorate, decoration={snake},thick},
  dashdot/.style={dash pattern=on .4pt off 3pt on 4pt off 3pt,thick}
}

 \newcommand{\finP}{\Pi^{(1)}_{\mathrm{F}}}

\begin{document}

\title{The $\beta$-function for Yukawa theory at large $N_f$}

\author{Tommi Alanne,}
\author{Simone Blasi}

\affiliation{Max-Planck-Institut f\"{u}r Kernphysik, Saupfercheckweg 1, 69117 Heidelberg, Germany}

\emailAdd{tommi.alanne@mpi-hd.mpg.de}
\emailAdd{simone.blasi@mpi-hd.mpg.de}

\abstract{
We compute the $\beta$-function for a massless
Yukawa theory in a closed form at the order $\mathcal{O}(1/N_f)$
in the spirit of the expansion in a 
large number of flavours $N_f$. We find an analytic expression with a finite radius of convergence, 
and the first singularity
occurs at the coupling value $K=5$.
}

\maketitle
\newpage

\section{Introduction}
The success of the Standard Model in describing the electroweak scale phenomena notwithstanding the apparent 
problems with the high-energy behaviour have lead to revival of interest in better understanding the UV 
properties of general 
gauge-Yukawa theories, see e.g. Refs~\cite{Litim:2014uca,Antipin:2017ebo,Eichhorn:2016esv}. 
In particular, gauge-Yukawa theories with a large number of fermion flavours, $N_f$, provide
interesting candidates within the asymptotic-safety framework as opposed to the traditional 
asymptotic-freedom paradigm~\cite{Gross:1973ju,Politzer:1973fx}.

The groundwork for these considerations was laid few decades ago with the computation of the leading large-$N_f$ behaviour
of the gauge $\beta$-functions~\cite{Espriu:1982pb,PalanquesMestre:1983zy,Gracey:1996he} for $N_f$ fermion charged under the 
gauge group; see also Refs~\cite{Holdom:2010qs,Shrock:2013cca}. 
The leading $1/N_f$ contribution to the $\beta$-function is obtained by 
resumming the gauge self-energy diagrams with ever increasing chain of fermion bubbles constituting a power series in 
$K=\alpha N_f/\pi$. It was noticed that this series has a finite radius of convergence; in the case of $\mathrm{U}(1)$ gauge group $K=15/2$.
Furthermore, the leading $1/N_f$ contribution to the $\mathrm{U}(1)$ $\beta$-function has a negative pole at $K=15/2$, 
thereby suggesting that this behaviour could cure the Landau-pole behaviour of the SM $\mathrm{U}(1)$ coupling, see e.g. Refs~\cite{Holdom:2010qs,Mann:2017wzh,Pelaggi:2017abg}.

Recently, a further step towards a  more complete understanding of these models was achieved by working out the 
leading $1/N_f$ contribution from the gauge sector to a Yukawa coupling~\cite{Kowalska:2017pkt}; 
an extension to semi-simple gauge groups was discussed in 
Ref.~\cite{Antipin:2018zdg}. However, only a single
fermion flavour was assumed to couple to the scalar, and the scalar self-energy remained uneffected by 
the $N_f$ fermion bubbles. Our work is the first step to bridge this remaining gap: we provide the leading $1/N_f$ $\beta$-function 
for pure Yukawa theory, where $N_f$ flavours of fermions couple to the scalar field via Yukawa interaction. We leave the 
more detailed study within a general gauge-Yukawa framework for future work. Interestingly, the pure Yukawa model is closely related to 
the Gross--Neveu--Yukawa model, whose critical exponents have been recently computed up to $1/N_f^2$~\cite{Gracey:2017fzu,Manashov:2017rrx}; 
see also the earlier studies on the Gross--Neveu model e.g. Refs~\cite{Gracey:1990wi,Vasiliev:1992wr}.

The paper is organized as follows: In Sec.~\ref{sec:def} we introduce the framework and notations and in 
Sec.~\ref{sec:renC} give the expressions for the renormalization constants. In Sec.~\ref{sec:resum} we perform the 
resummations of the bubble chains and give closed form expressions for the renormalization constants. In 
Sec.~\ref{sec:beta} we collect the results, and write down the final expression for the $\beta$-function, and  
in Sec.~\ref{sec:concl} we conclude. 
Explicit formulas for the loop integrals are given in Appendix~\ref{sec:loops}.

\section{The framework and definitions}
\label{sec:def}
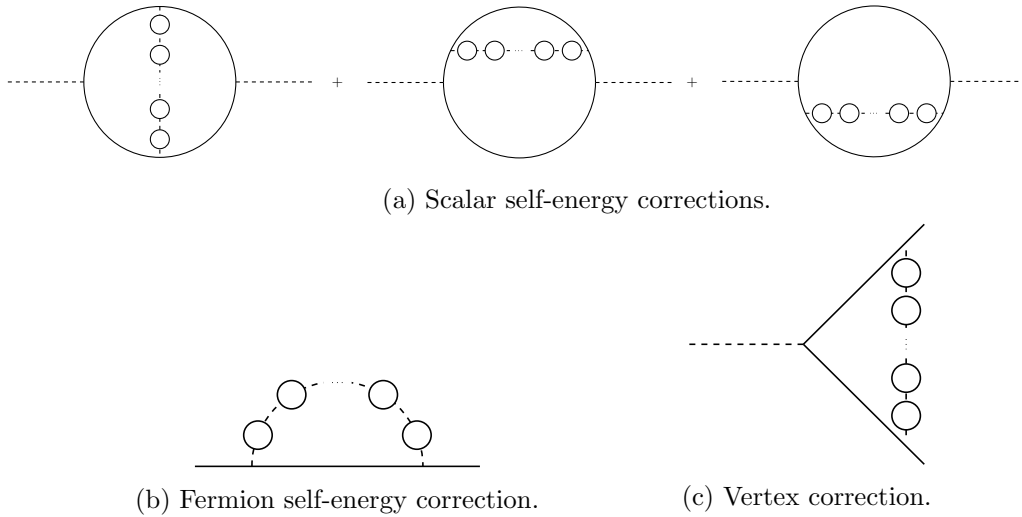
\begin{figure}
\centering
\begin{subfigure}{\textwidth}
    \scalebox{0.5}{
\begin{minipage}[c]{.25\textwidth}
  \begin{tikzpicture}[node distance=2cm]
    \coordinate (v1);
    \coordinate[right = of v1] (v2);
    \coordinate[right = of v2] (v3);
    \coordinate[right = of v3] (v4);
    \coordinate[right = of v4] (v5);
    \draw[rscalar] (v1) -- (v2);
    \draw[mfermion] (v2) arc(180:90:2) coordinate (v8) arc(90:0:2);
    \draw[mfermion] (v2) arc(-180:-90:2) coordinate (v9) arc(-90:0:2);
    \draw[rscalar] (v4) -- (v5);
    \draw[rscalar] (v8) --  (v9) 
	node[pos=0.12,draw,solid,whiteblob,minimum size=0.5 cm] {}
	node[pos=0.32,draw,solid,whiteblob,minimum size=0.5 cm] {}
	node[pos=0.44] (v11) {}
	node[pos=0.56] (v12) {}
	node[midway,circle,fill=white,dotted, minimum size=0.5 cm]{}
	node[pos=0.68,draw,solid,whiteblob,minimum size=0.5 cm] {}
	node[pos=0.88,draw,solid,whiteblob,minimum size=0.5 cm] {};
    \draw[dotted] (v11)--(v12);
  \end{tikzpicture}
\end{minipage}
  ~\hspace{4.5cm}$+$\hspace{0.5cm}
\begin{minipage}[c]{.25\textwidth}
    \vspace{-.25cm}
  \begin{tikzpicture}[node distance=2cm]
    \coordinate (v1);
    \coordinate[right = of v1] (v2);
    \coordinate[right = of v2] (v3);
    \coordinate[right = of v3] (v4);
    \coordinate[right = of v4] (v5);
    \draw[rscalar] (v1) -- (v2);
    \draw[mfermion] (v2) arc(180:155:2) coordinate (v8) arc(155:25:2) coordinate (v9) arc(25:0:2);
    \draw[mfermion] (v2) arc(-180:0:2);
    \draw[rscalar] (v4) -- (v5);
    \draw[rscalar] (v8) --  (v9) 
	node[pos=0.12,draw,solid,whiteblob,minimum size=0.5 cm] {}
	node[pos=0.32,draw,solid,whiteblob,minimum size=0.5 cm] {}
	node[pos=0.43] (v11) {}
	node[pos=0.57] (v12) {}
	node[midway,circle,fill=white,dotted, minimum size=0.5 cm]{}
	node[pos=0.68,draw,solid,whiteblob,minimum size=0.5 cm] {}
	node[pos=0.88,draw,solid,whiteblob,minimum size=0.5 cm] {};
    \draw[dotted] (v11)--(v12);
  \end{tikzpicture}
\end{minipage}
  \hspace{4.5cm}$+$ \hspace{.5cm}\vspace{2cm}
\begin{minipage}[c]{.25\textwidth}
\vspace{0.25cm}
  \begin{tikzpicture}[node distance=2cm]
    \coordinate (v1);
    \coordinate[right = of v1] (v2);
    \coordinate[right = of v2] (v3);
    \coordinate[right = of v3] (v4);
    \coordinate[right = of v4] (v5);
    \draw[rscalar] (v1) -- (v2);
    \draw[mfermion] (v2) arc(-180:-155:2) coordinate (v8) arc(-155:-25:2) coordinate (v9) arc(-25:0:2);
    \draw[mfermion] (v2) arc(180:0:2);
    \draw[rscalar] (v4) -- (v5);
    \draw[rscalar] (v8) --  (v9) 
	node[pos=0.12,draw,solid,whiteblob,minimum size=0.5 cm] {}
	node[pos=0.32,draw,solid,whiteblob,minimum size=0.5 cm] {}
	node[pos=0.43] (v11) {}
	node[pos=0.57] (v12) {}
	node[midway,circle,fill=white,dotted, minimum size=0.5 cm]{}
	node[pos=0.68,draw,solid,whiteblob,minimum size=0.5 cm] {}
	node[pos=0.88,draw,solid,whiteblob,minimum size=0.5 cm] {};
    \draw[dotted] (v11)--(v12);
  \end{tikzpicture}
\end{minipage}
  }
  \caption{Scalar self-energy corrections.}
  \label{fig:sBubbleChain}
\end{subfigure}\\
\begin{subfigure}{0.4\textwidth}\quad
\centering
~\\
\vspace{1.9cm}
    \scalebox{0.75}{
  \begin{tikzpicture}[node distance=1cm]
    \coordinate (v1);
    \coordinate[right = of v1] (v2);
    \coordinate[right = of v2] (v31);
    \coordinate[right = of v31] (v32);
    \coordinate[right = of v32] (v3);
    \coordinate[right = of v3] (v4);
    \draw[mfermion] (v1) -- (v4);
    \draw[rscalar] (v2) arc(180:0:1.5) 
	node[pos=0.12,draw,solid,whiteblob,minimum size=0.5 cm] {}
	node[pos=0.32,draw,solid,whiteblob,minimum size=0.5 cm] {}
	node[pos=0.45] (v11) {}
	node[pos=0.55] (v12) {}
	node[midway,circle,fill=white,dotted, minimum size=0.5 cm]{}
	node[pos=0.68,draw,solid,whiteblob,minimum size=0.5 cm] {}
	node[pos=0.88,draw,solid,whiteblob,minimum size=0.5 cm] {};
    \draw[dotted] (v11)--(v12);
  \end{tikzpicture}
  }
  \caption{Fermion self-energy correction.}
  \label{fig:fBubbleChain}
\end{subfigure}
\vspace{2cm}
\begin{subfigure}{0.4\textwidth}
\centering
    \scalebox{0.75}{
  \begin{tikzpicture}[node distance=1cm]
    \coordinate (v1);
    \coordinate[right = of v1] (v21);
    \coordinate[right = of v21] (v2);
    \coordinate[above right = of v2] (v31);
    \coordinate[above right = of v31] (v32);
    \coordinate[above right = of v32] (v3);
    \coordinate[below right = of v2] (v41);
    \coordinate[below right = of v41] (v42);
    \coordinate[below right = of v42] (v4);
    \draw[rscalar] (v1) -- (v2);
    \draw[mfermion] (v2) -- (v3) node[pos=0.85] (v5) {};
    \draw[mfermion] (v2) -- (v4) node[pos=0.85] (v6) {};
    \draw[rscalar] (v5) --  (v6) 
	node[pos=0.12,draw,solid,whiteblob,minimum size=0.5 cm] {}
	node[pos=0.32,draw,solid,whiteblob,minimum size=0.5 cm] {}
	node[pos=0.43] (v11) {}
	node[pos=0.57] (v12) {}
	node[midway,circle,fill=white,dotted, minimum size=0.5 cm]{}
	node[pos=0.68,draw,solid,whiteblob,minimum size=0.5 cm] {}
	node[pos=0.88,draw,solid,whiteblob,minimum size=0.5 cm] {};
    \draw[dotted] (v11)--(v12);
  \end{tikzpicture}
  }
  \caption{Vertex correction.}
  \label{fig:vBubbleChain}
\end{subfigure}
\vspace{-1.5cm}
\caption{Scalar self-energy, fermion self-energy, and vertex corrections due to a chain of fermion bubbles.}
\label{fig:bubbleChains}
\end{figure}

We consider the massless Yukawa theory for a real scalar field, $\phi$,
and a fermionic multiplet, $\psi$, consisting of $N_f$ flavours 
interacting through the usual Yukawa interaction:
\begin{equation}
 \mathcal{L}_{\mathrm{Yuk}} =  g  \bar{\psi} \psi \phi.
\end{equation}
We define the rescaled coupling,  
\begin{equation}
 K \equiv \frac{g^2}{4 \pi^2} N_f,
\end{equation}
which is kept constant in the limit $N_f\to\infty$.
The $\beta$-function of the rescaled coupling, $K$, can then be expanded in powers of $1/N_f$ as
\begin{equation}\label{eq:F0F1}
 \beta(K) \equiv \frac{\mathrm{d}K}{\mathrm{d}\ln\mu}= K^2\left[ F_0 + \frac{1}{N_f} F_1(K)\right] + \mathcal{O}\left(1/N_f^2\right).
\end{equation}
The purpose of this paper is to compute $F_0$ and $F_1(K)$.
The former is entirely fixed at the one-loop level and can be
derived just by rescaling the well-known 
result for the $\beta$-function at that order,
while the evaluation of $F_1(K)$ requires the resummation of diagrams in Fig.~\ref{fig:bubbleChains}
involving all-order fermion-bubble chains.

The $\beta$-function can be obtained from 
\begin{equation}\label{eq:beta}
 \beta = K^2 \frac{\partial G_1(K)}{\partial K},
\end{equation}
where $G_1$ is defined by 
\begin{equation}\label{eq:ZK}
\text{ln}\,Z_K \equiv \text{ln}\, (Z_S^{-1} Z_F^{-2} Z_V^2) = \sum_{n=1}^\infty \frac{G_n(K)}{\epsilon^n},
\end{equation}
and $Z_S$, $Z_F$, and $Z_V$ are the renormalization constants for the scalar
wave function, the fermion wave function, and the 1PI vertex,
respectively.
The scalar wave function renormalization constant is determined via
\begin{equation}\label{eq:Z_Sdef}
Z_S = 1 - \text{div} \{ Z_S \Pi_0(p^2, Z_K K,\epsilon) \}, 
\end{equation}
where $\Pi_0(p^2,K_0,\epsilon)$ is the scalar self-energy divided by $p^2$,  where $p$ is the external momentum. 
Here and in the following, 
$\text{div}{X}$ denotes the poles of $X$ in $\epsilon$. 
The self-energy can be written as 
\begin{equation}\label{eq:PI0}
 \Pi_0(p^2,K_0,\epsilon) =
 K_0 \Pi^{(1)}(p^2,\epsilon) + \frac{1}{N_f} \sum_{n=2}^\infty K_0^n \Pi^{(n)}(p^2,\epsilon),
\end{equation}
where $\Pi^{(1)}$ gives the one-loop result, and $\Pi^{(n)}$ the $n$-loop part containing $n-2$ fermion bubbles in the chain, 
and summing over the topologies given in Fig.~\ref{fig:sBubbleChain}. Other contributions are of higher order 
in $1/N_f$ and are thus omitted.

For the fermion self-energy and vertex renormalization constants, the lowest non-trivial contributions are
already $\mathcal{O}(1/N_f)$, and we, therefore, have
\begin{equation}\label{eq:Z_Fdef}
 Z_F = 1 - \text{div} \left\{ \Sigma_0(p^2, Z_K K, \epsilon) \right\},
\end{equation}
\begin{equation}
 \Sigma_0(p^2,K_0,\epsilon) = \frac{1}{N_f} \sum_{n=1}^\infty K_0^n \Sigma^{(n)}(p^2,\epsilon),
\end{equation}
where $\Sigma^{(n)}$ is depicted in Fig.~\ref{fig:fBubbleChain} with $n-1$ fermion bubbles. Similarly,
\begin{equation}\label{eq:Z_Vdef}
 Z_V = 1 - \text{div} \left\{ V_0(p^2, Z_K K, \epsilon) \right\},
\end{equation}
\begin{equation}
V_0(p^2,K_0,\epsilon) = \frac{1}{N_f} \sum_{n=1}^\infty K_0^n V^{(n)}(p^2,\epsilon),
\end{equation}
where $V^{(n)}$ again contains $n-1$ fermion bubbles and is shown diagrammatically in Fig~\ref{fig:vBubbleChain}. 

Finally, we briefly comment on the scalar 
three-point and four-point functions,
assuming that they are generated
via fermion loops: the former exactly vanishes for massless
fermions, while the latter is found to be already
$\mathcal{O}(1/N_f)$ at the lowest order. Therefore, they
can be neglected for the purpose of our analysis.

\section{Renormalization constants}
\label{sec:renC}
In this section our goal is to extract the contributions to the renormalization constants that are
$\mathcal{O}(1/N_F)$ and relevant for the computation of the $\beta$-function.

Our starting point for $Z_S$ is Eq.~\eqref{eq:Z_Sdef}. Using the expansion of the scalar self-energy, Eq.~\eqref{eq:PI0},
we obtain
\begin{equation}\label{eq:Z_S2}
\begin{split}
 Z_S & = 1 - \text{div} \bigg\{ Z_S Z_K K \Pi^{(1)}(p^2,\epsilon) + 
\frac{1}{N_f} \sum_{n=2}^\infty Z_S ( Z_K K)^n \Pi^{(n)}(p^2,\epsilon) \bigg\}. 
 \end{split}
\end{equation}
Recalling that $Z_K \equiv Z_S^{-1}Z_F^{-2}Z_V^2$ and substituting Eqs~\eqref{eq:Z_Fdef} and~\eqref{eq:Z_Vdef}, 
the first term between brackets can be written as
\begin{multline}
\text{div} \left \{ Z_S Z_K \Pi^{(1)}(p^2,\epsilon) K \right\} \\
=K \text{div}\left\{ \Pi^{(1)}\right\}
+ \frac{1}{N_f} \text{div} \left\{ 2 K \, \text{div}\left\{\Sigma_0(p^2,Z_K K,\epsilon) - 
V_0(p^2,Z_K K,\epsilon) \right\} 
\Pi^{(1)}(p^2,\epsilon) \right\}. 
\end{multline}
The $\Pi^{(1)}$ part corresponds to the one-loop diagram and is given by
\begin{equation}
\begin{split}
 \Pi^{(1)}(p^2,\epsilon) \equiv &
 \text{div}\left\{\Pi^{(1)}\right\} + \finP(p^2,\epsilon) =
 \frac{1}{(4\pi)^{d/2-2}} \frac{G(1,1)}{2} (-p^2)^{d/2-2}  \\
 =& \frac{1}{\epsilon} 
 + \finP(p^2,\epsilon),
 \end{split}
\end{equation}
where $d=4-\epsilon$, the loop function, $G(1,1)$, is defined in Eq.~\eqref{eq:G2} in Appendix~\ref{sec:app1},
and we have introduced the notation $\finP$ to indicate the finite part of $\Pi^{(1)}$. 
Then, 
\begin{equation}\begin{split}\label{eq:Z_Sfirst}
\text{div} &\left \{ Z_S  Z_K \Pi^{(1)}(p^2,\epsilon) K \right\} \\
&=  \frac{K}{\epsilon}
+ \frac{1}{N_f} \text{div} \bigg\{ 2 K \, \text{div}\left\{\Sigma_0(p^2,Z_K K,\epsilon) - 
V_0(p^2,Z_K K,\epsilon) \right\} \\ &
\qquad\qquad\qquad\quad\times 
\left( \text{div}\left\{\Pi^{(1)}\right\} + \finP(p^2,\epsilon)\right) \bigg\} \\ 
&=  \frac{K}{\epsilon} +
\frac{1}{N_f} \text{div} \left\{ 2 K \finP(p^2,\epsilon)
\left[\Sigma_0(p^2,Z_K K,\epsilon) - V_0(p^2,Z_K K,\epsilon) \right] \right\} \\
&\quad + \frac{1}{N_f}\times\ \text{higher poles},
\end{split}
\end{equation}
where the higher poles, i.e., higher than $1/\epsilon$,
arise from the product of two divergent parts and will be omitted
because they play no role in what follows. 
Then, at the lowest order in $1/N_f$,  
\begin{equation}
 Z_S = 1 - \frac{K}{\epsilon} + \mathcal{O}\left(1/N_f \right).
\end{equation}
Therefore, every time $Z_K K$ appears in the argument of $\Sigma_0$ and
$V_0$, it can be replaced by $ K\left(1 - \frac{K}{\epsilon}\right)^{-1}$; 
the additional contributions are higher order in $1/N_f$.
For Eq.~\eqref{eq:Z_Sfirst}, we arrive at
\begin{multline}
 \text{div} \left \{ Z_S Z_K \Pi^{(1)}(p^2,\epsilon) K \right\}\\
 = \frac{K}{\epsilon}
+ \sum_{n=1}^\infty K^{n+1} \text{div} \left\{ 2  \finP(p^2,\epsilon) 
 \left(1-\frac{K}{\epsilon}\right)^{-n} \left[ \Sigma^{(n)}(p^2,\epsilon) -
 V^{(n)}(p^2,\epsilon) \right]
\right\}. 
\end{multline}
Similarly, the second term of Eq.~\eqref{eq:Z_S2} reads
\begin{equation}
 \frac{1}{N_f} \text{div} \left\{ \sum_{n=2}^\infty Z_S ( Z_S^{-1} K)^n \Pi^{(n)}(p^2,\epsilon) 
 \right\} =  \frac{1}{N_f} \sum_{n=2}^\infty K^n \text{div}\left\{ 
 \left( 1 - \frac{K}{\epsilon} \right)^{1-n} \Pi^{(n)}(p^2,\epsilon) \right\}.
\end{equation}
Altogether, we can write $Z_S$ as
\begin{multline}
 Z_S = 1 - \frac{K}{\epsilon} - \frac{1}{N_f} \sum_{n=2}^\infty K^n 
 \left\{ \left( 1 - \frac{K}{\epsilon} \right)^{1-n}
 \left( 2  \finP
 \left[ \Sigma^{(n-1)}
 - V^{(n-1)} \right] + \Pi^{(n)}
 \right)
 \right\},
\end{multline}
where the explicit functional dependence on $(p^2,\epsilon)$ has been omitted to lighten the notation.
Using the binomial expansion,
\begin{equation}
 \left(1 - \frac{K}{\epsilon}\right)^{1-n} =
 \sum_{i=0}^\infty \left( \begin{array}{c} n + i - 2 \\ i \end{array} \right)
 \frac{K^i}{\epsilon^i}
\end{equation}
and performing a shift in the summation, $n \rightarrow n - i$,
we find our final expression for $Z_S$:
\begin{equation}\label{eq:Z_Sfinal}
 Z_S = 1 - \frac{K}{\epsilon} 
   - \frac{1}{N_f} \sum_{n=2}^\infty K^n \text{div} \left\{
   \sum_{i=0}^{n-2}
  \left( \begin{array}{c} n-2 \\ i \end{array} \right)
  \frac{1}{\epsilon^i}
  \left( 2 \finP \left( \Sigma^{(n-i-1)}-V^{(n-i-1)} 
  \right) + \Pi^{(n-i)} 
  \right) \right\}. 
\end{equation}
We notice that Eq.~\eqref{eq:Z_Sfinal} differs essentially 
from its counterpart in
the QED~\cite{PalanquesMestre:1983zy} 
because of the contribution from the fermion self-energy
and the vertex, which exactly cancel in QED because of the
Ward identity.

The expression for $Z_F$ can be derived from Eq.~\eqref{eq:Z_Fdef} in a similar 
manner:
\begin{equation}\label{eq:Z_Ffinal}
\begin{split}
Z_F & = 1 - \frac{1}{N_f} \sum_{n=1}^\infty \text{div}\left\{
\left( Z_K K \right)^n \Sigma^{(n)}(p^2,\epsilon) \right\} \\
& = 1 - \frac{1}{N_f} \sum_{n=1}^\infty K^n 
\text{div}\left\{\left(1-\frac{K}{\epsilon}\right)^{-n}\Sigma^{(n)}(p^2,\epsilon)
\right\} \\
& = 1 - \frac{1}{N_f} \sum_{n=1}^\infty K^n
\text{div} \left\{ \sum_{i=0}^{n-1} 
\left( \begin{array}{c} n-1 \\ i \end{array} \right) \frac{1}{\epsilon^i} 
\Sigma^{(n-i)}(p^2,\epsilon) \right\},
\end{split}
\end{equation}
where we have again performed the same shift $n \rightarrow n-i$ in the last line.
The derivation of $Z_V$ is completely analogous, and we can readily write the expression for $Z_V$:
\begin{equation}\label{eq:Z_Vfinal}
Z_V =  1 - \frac{1}{N_f} \sum_{n=1}^\infty K^n
\text{div} \left\{ \sum_{i=0}^{n-1} 
\left( \begin{array}{c} n-1 \\ i \end{array} \right) \frac{1}{\epsilon^i} 
V^{(n-i)}(p^2,\epsilon) \right\}.
\end{equation}

\section{Resummation}
\label{sec:resum}

In this section we provide closed formulas for Eqs~\eqref{eq:Z_Sfinal},
\eqref{eq:Z_Ffinal}, and~\eqref{eq:Z_Vfinal}.

\subsection{The vertex}
By explicit computation, the $n$-loop contribution to $V_0$ is
\begin{equation}\label{eq:Vn}
\begin{split}
 V^{(n)}(p^2,\epsilon) =& \frac{(-1)^{n}}{4} 
  \left(\frac{1}{(4 \pi)^{d/2-2}} \right)^n 
 \left(\frac{G(1,1)}{2} \right)^{n-1} (-p^2)^{n(d/2-2)} \\
  &\times G\left(1,1-(n-1)(d/2-2) \right),
\end{split}
\end{equation}
where $G(n_1,n_2)$ is defined in Eq.~\eqref{eq:G2}.
We notice that, as in Ref.~\cite{PalanquesMestre:1983zy}, Eq.~\eqref{eq:Vn}
allows for the following expansion:
\begin{equation}\label{eq:V^nexp}
V^{(n)}(p^2,\epsilon) = (-1)^{n} \frac{1}{n \epsilon^n} \frac{v(p^2,\epsilon,n)}{2},
\end{equation}
where 
\begin{equation}
 v(p^2,\epsilon,n) = \sum_{j=0}^\infty v_j(p^2,\epsilon) (n \epsilon)^j,
\end{equation}
and $v_j(p^2,\epsilon)$ are regular in the limit $\epsilon \rightarrow 0$ for all $j$.
In particular, $v_0(\epsilon)$ is independent of $p^2$ and is explicitly given by
\begin{equation}\label{eq:v0def}
v_0(\epsilon) = \frac{2 \Gamma(2-\epsilon)}
{\Gamma\left(1-\frac{\epsilon}{2}\right)^2
\Gamma \left(2 - \frac{\epsilon}{2} \right) 
\Gamma \left( \frac{\epsilon}{2} \right)\,\epsilon}.
\end{equation}
Substituting Eqs~\eqref{eq:Vn} and~\eqref{eq:V^nexp} in Eq.~\eqref{eq:Z_Vfinal}, we find:
\begin{equation}\label{eq:Z_Vsum}
Z_V = 1 - \frac{1}{N_f} \sum_{n=1}^\infty (-K)^n\text{div}\left\{
\sum_{j=0}^{n-1} \frac{1}{\epsilon^{n-j}} \sum_{i=0}^{n-1} 
\left( \begin{array}{c} n-1 \\ i \end{array} \right)(-1)^i
(n-i)^{j-1} \frac{v_j(p^2,\epsilon)}{2}
\right\}.
\end{equation}
Then, by using the result of Ref.~\cite{PalanquesMestre:1983zy},
\begin{equation}
 \sum_{i=0}^{n-1}\left( \begin{array}{c} n-1 \\ i \end{array} \right)(-1)^i
(n-i)^{j-1} = - \delta_{j,0} \frac{(-1)^n}{n}, \ j=0,\dots,n-1,
\end{equation}
Eq.~\eqref{eq:Z_Vsum} gets simplified to
\begin{equation}\label{eq:Z_Vsim}
 Z_V = 1 + \frac{1}{2 N_f} \sum_{n=1}^\infty \frac{K^n}{\epsilon^n} \frac{v_0(\epsilon)}{n }.
\end{equation}
Expanding $v_0(\epsilon)$ as 
\begin{equation}
 v_0(\epsilon) = \sum_{j=0}^\infty v_0^{(j)} \epsilon^j
\end{equation}
and keeping only the $1/\epsilon$ pole of Eq.~\eqref{eq:Z_Vsim}, we find
the closed formula for $Z_V$:
\begin{equation}
 Z_V  = 1 +  \frac{1}{2 \epsilon N_f} \sum_{n=1}^\infty
\frac{K^n}{n}v_0^{(n-1)} = 1 + \frac{1}{2 \epsilon N_f}  \int_0^K v_0(t) \mathrm{d}t.
\end{equation}

\subsection{The fermion self-energy}
The $n$-loop contribution to $\Sigma_0$ is found to be
\begin{equation}\label{eq:Sigmaexp}
\begin{split}
 \Sigma^{(n)}(p^2,\epsilon) =&
 - \frac{(-1)^{n}}{8}
 \left( \frac{1}{(4 \pi)^{d/2-2}} \right)^n  
 \left( \frac{G(1,1)}{2} \right)^{n-1} (-p^2)^{n(d/2-2)} \\ 
 &\times
 \left[ G(1,1-(n-1)(d/2-2))-G(1,-(n-1)(d/2-2)) \right].
\end{split}
\end{equation}
Similarly to Eq.~\eqref{eq:Vn}, Eq.~\eqref{eq:Sigmaexp} can be expanded as
\begin{equation}
\Sigma^{(n)}(p^2,\epsilon)= - (-1)^{n} \frac{1}{n \epsilon^n} 
\frac{\sigma(p^2,\epsilon,n)}{4},
\end{equation}
where 
\begin{equation}
 \sigma(n,\epsilon,p^2) = \sum_{j=0}^{\infty} \sigma_j(p^2,\epsilon) (n \epsilon)^j,
\end{equation}
and $\sigma_j(p^2,\epsilon)$ are regular for $\epsilon \rightarrow 0$. 
Again, $\sigma_0(\epsilon)$ is independent of $p^2$, and it is given by
\begin{equation}
\sigma_0(\epsilon) = - \frac{ 2^{5 - \epsilon} 
\Gamma\left( \frac{3}{2} - \frac{\epsilon}{2} \right)}
{ \sqrt{\pi}(4 - \epsilon) \Gamma\left(-\frac{\epsilon}{2}\right) \epsilon}
\frac{\text{sin}\left(\frac{\pi \epsilon}{2}\right)}{\pi \epsilon}.
\end{equation}
Using the same procedure as in the previous section, we find that only $\sigma_0(\epsilon)$
contributes to $Z_F$. Keeping only the $1/\epsilon$ pole, the closed formula for
$Z_F$ is
\begin{equation}
 Z_F = 1 - \frac{1}{4 \epsilon N_f} \int_0^K \sigma_0(t) \mathrm{d}t.
\end{equation}

\subsection{The scalar self-energy}
The evaluation of the bubble diagrams in Fig.~\ref{fig:sBubbleChain} is quite
cumbersome and is discussed in Appendix~\ref{sec:app2}.
Here, we notice that the expression for $\Pi^{(n)}(p^2,\epsilon)$, $n \geq 2$, allows
for the following expansion:
\begin{equation}
 \Pi^{(n)}= - \frac{3}{2} \frac{(-1)^n}{n (n-1) \epsilon^n}
 \pi(p^2,\epsilon,n),
\end{equation}
where
\begin{equation}
 \pi(p^2,\epsilon,n) = \sum_{j=0}^\infty \pi_j(p^2,\epsilon)(n \epsilon)^j,
\end{equation}
and $\pi_j(p^2,\epsilon)$ are regular for $\epsilon \rightarrow 0$.
Similarly to the previous cases, $\pi_0(\epsilon)$ is independent of $p^2$. 

In view of Eq.~\eqref{eq:Z_Sfinal}, we define
\begin{equation}
 2 \finP(p^2,\epsilon) \left( \Sigma^{(n-1)}(p^2,\epsilon) -V^{(n-1)}(p^2,\epsilon)  
  \right) + \Pi^{(n)}(p^2,\epsilon)  
  \equiv \frac{(-1)^n}{n (n-1) \epsilon^n} \xi(p^2,\epsilon,n), 
\end{equation}
where
\begin{equation}\label{eq:csindef}
\xi(p^2,\epsilon,n) \equiv n \epsilon \, \finP
\left( \frac{\sigma(p^2,\epsilon,n-1)}{2}+ v(p^2,\epsilon,n-1) \right)
- \frac{3}{2} \pi(p^2,\epsilon,n),
\end{equation}
and
\begin{equation}
\xi(p^2,\epsilon,n) = \sum_{j=0}^\infty \xi_j(p^2,\epsilon) (n \epsilon)^j,
\end{equation}
with $\xi_j(\epsilon,p^2)$ regular for $\epsilon \rightarrow 0$ for all $j$.
In particular, $\xi_0(\epsilon)$ is independent of $p^2$ and is explicitly given by
\begin{equation}
\xi_0(\epsilon)=-\frac{(1-\epsilon)\Gamma(4-\epsilon)}{\Gamma\left(2-\frac{\epsilon}{2}\right)
    \Gamma\left(3-\frac{\epsilon}{2}\right)\pi\epsilon}\sin\left(\frac{\pi\epsilon}{2}\right)
\end{equation}
Then, using the above definitions, Eq.~\eqref{eq:Z_Sfinal} can be written as
\begin{equation}
\begin{split}
Z_S & = 1 - \frac{K}{\epsilon} - \frac{1}{N_f} \sum_{n=2}^\infty K^n \text{div} \left\{
   \sum_{i=0}^{n-2}
  \left( \begin{array}{c} n-2 \\ i \end{array} \right)
  \frac{1}{\epsilon^i} 
  \frac{(-1)^{n-i}}{(n-i)(n-i-1)\epsilon^{n-i}}
  \xi(p^2,\epsilon,n-i)
  \right\} \\
  & = 1 - \frac{K}{\epsilon} - \frac{1}{N_f} \sum_{n=2}^\infty (-K)^n
  \text{div} \left\{\sum_{j=0}^{n-1}
\frac{1}{\epsilon^{n-j}}   \xi_j(p^2,\epsilon)
   \sum_{i=0}^{n-2}
  \left( \begin{array}{c} n-2 \\ i \end{array} \right)(-1)^{i}
  \frac{(n-i)^{j-1}}{(n-i-1)}
  \right\}.
  \end{split}
  \end{equation}
Moreover, we find that
\begin{equation}
 \sum_{i=0}^{n-2}
  \left( \begin{array}{c} n-2 \\ i \end{array} \right) (-1)^{i}
  \frac{(n-i)^{j-1}}{(n-i-1)} = \begin{cases} \frac{(-1)^n}{n} & j = 0 \\
  \frac{(-1)^n}{n-1} & j = 1,\dots,n-1 \end{cases},
\end{equation}
and therefore the expression for $Z_S$ can be significantly simplified:
\begin{equation}
\label{eq:ZSxi}
\begin{split}
Z_S & = 1 - \frac{K}{\epsilon} - \frac{1}{N_f} \sum_{n=2}^\infty K^n \, \text{div} \left\{
\frac{1}{\epsilon^n}
   \left( \frac{\xi_0(\epsilon)}{n} + 
   \frac{1}{n-1} \sum_{j=1}^{n-1} \xi_j(p^2,\epsilon) \epsilon^j \right)
   \right\} \\
& = 1 - \frac{K}{\epsilon} - \frac{1}{N_f} \sum_{n=2}^\infty K^n \, \text{div} \left\{
\frac{1}{\epsilon^n}
   \left( \frac{\xi_0(\epsilon)}{n} + 
   \frac{1}{n-1} \sum_{j=1}^{\infty} \xi_j(p^2,\epsilon) \epsilon^j \right)
   \right\}  \\   
& = 1 - \frac{K}{\epsilon} - \frac{1}{N_f} \sum_{n=2}^\infty K^n \, \text{div} \left\{
\frac{1}{\epsilon^n}
   \left( \frac{\xi_0(\epsilon)}{n} + 
  \frac{\xi(p^2,\epsilon,1) - \xi_0(\epsilon)}{n-1} \right)
   \right\}, 
   \end{split}
\end{equation}
where in the second line we extended the sum over $j$ up to $\infty$ without
affecting the result, since all the terms for $ j > n-1$ are finite.
The function $\xi(p^2,\epsilon,1)$, corresponding to 
\begin{equation}
 \xi(p^2,\epsilon,1) \equiv \sum_{j=0}^\infty \xi_j(p^2,\epsilon)\epsilon^j,
\end{equation}
can be evaluated by taking in Eq.~\eqref{eq:csindef}  the limit 
$n\rightarrow 1$, although the latter is formally
defined for $n \geq 2$. We find the following expression:
\begin{equation}\label{eq:cancel}
\xi(p^2,\epsilon,1) =
-\frac{\Gamma(4-\epsilon)}{\Gamma\left(2-\frac{\epsilon}{2}\right)
    \Gamma\left(3-\frac{\epsilon}{2}\right)\pi\epsilon}\sin\left(\frac{\pi\epsilon}{2}\right)
\equiv \xi(\epsilon,1).
\end{equation}

Few comments are in order: Eq.~\eqref{eq:cancel}
ensures that $Z_S$ is independent of the external
momentum $p^2$, as it should. This result comes from
an exact cancellation among the different contributions of the scalar
self-energy, the fermion self-energy, and the vertex in Eq.~\eqref{eq:csindef}. 
In particular, we find that
\begin{equation}\label{eq:pi1}
\begin{split}
  \pi(p^2,\epsilon,1) & =  \frac{2}{3} \left(\frac{\sigma(p^2,\epsilon,0)}{2} + 
 v(p^2,\epsilon,0) \right)
 \left[1 + 1 \cdot \epsilon \, \finP(p^2,\epsilon) \right] \\
& =  \frac{2}{3} \left(\frac{\sigma_0(\epsilon)}{2} + v_0(\epsilon) \right)
 \left[1 + \epsilon \, \finP(p^2,\epsilon) \right],
 \end{split}
\end{equation}
and therefore
\begin{equation}\label{eq:xieps1}
 \xi(\epsilon,1) = - \frac{\sigma_0(\epsilon)}{2} - v_0(\epsilon),
\end{equation}
which is equivalent to Eq.~\eqref{eq:cancel}.
Interestingly, Eq.~\eqref{eq:pi1} only holds for $n=1$.
All in all, the $p^2$ independence of 
Eq.~\eqref{eq:cancel} provides a non-trivial check for our computation.
Moreover, we see that
\begin{equation}
 \xi_0(\epsilon) = (1-\epsilon) \xi(\epsilon,1).
\end{equation}

We are now ready to resum the series in Eq.~\eqref{eq:ZSxi}.
By expanding $\xi_0(\epsilon)$ as
\begin{equation}
 \xi_0(\epsilon) = \sum_{j=0}^\infty \xi_0^{(j)} \epsilon^j,
\end{equation}
the $\frac{1}{n}$ term in Eq.~\eqref{eq:ZSxi} is given by
\begin{equation}\begin{split}
\sum_{n=2}^\infty \frac{K^n}{\epsilon^n} \frac{\xi_0(\epsilon)}{n} 
& = \frac{1}{\epsilon} \sum_{n=2}^\infty \frac{K^n}{\epsilon^n}
\frac{\xi_0^{(n-1)}}{n} + \text{higher poles} \\
& =
\frac{1}{\epsilon} \left( \, \sum_{n=0}^\infty K^{n+1} \frac{\xi_0^{(n)}}{n+1}
- K \xi_0^{(0)} \right) + \text{higher poles} \\
& =  \frac{1}{\epsilon} \int_0^K \left[ \xi_0(t) - \xi_0(0) \right] \mathrm{d}t
+ \text{higher poles} .
\end{split}
\end{equation}
As for the $\frac{1}{n-1}$ term,
using $\xi_0(\epsilon)=(1-\epsilon) \xi(\epsilon,1)$ and expanding 
$\xi(\epsilon,1)$ as
\begin{equation}
 \xi(\epsilon,1) = \sum_{j=0}^\infty \tilde{\xi}^{(n)} \epsilon^j,
\end{equation}
we find
\begin{equation}
\begin{split}
 \sum_{n=2}^\infty \frac{K^n}{\epsilon^{n}}
 \frac{ \epsilon \xi(\epsilon,1)}{n-1}  & =
 \frac{K}{\epsilon} \sum_{n=0}^\infty \frac{K^{n+1}}{n+1}
  \tilde{\xi}^{(n)} + \text{higher poles} \\
& = \frac{K}{\epsilon} \int_0^K \xi(t,1)\mathrm{d}t
+ \text{higher poles}.
 \end{split}
\end{equation}
Finally, the closed formula for $Z_S$ reads
\begin{equation}
 Z_S = 1 - \frac{K}{\epsilon} - \frac{1}{ \epsilon N_f} 
 \int_0^K \left[ \xi_0(t) - \xi_0(0) + \xi(t,1) K \right] \mathrm{d}t.
\end{equation}

\section{The $\beta$-function}
\label{sec:beta}
\begin{figure}[t]
\centering
\includegraphics[width=8cm]{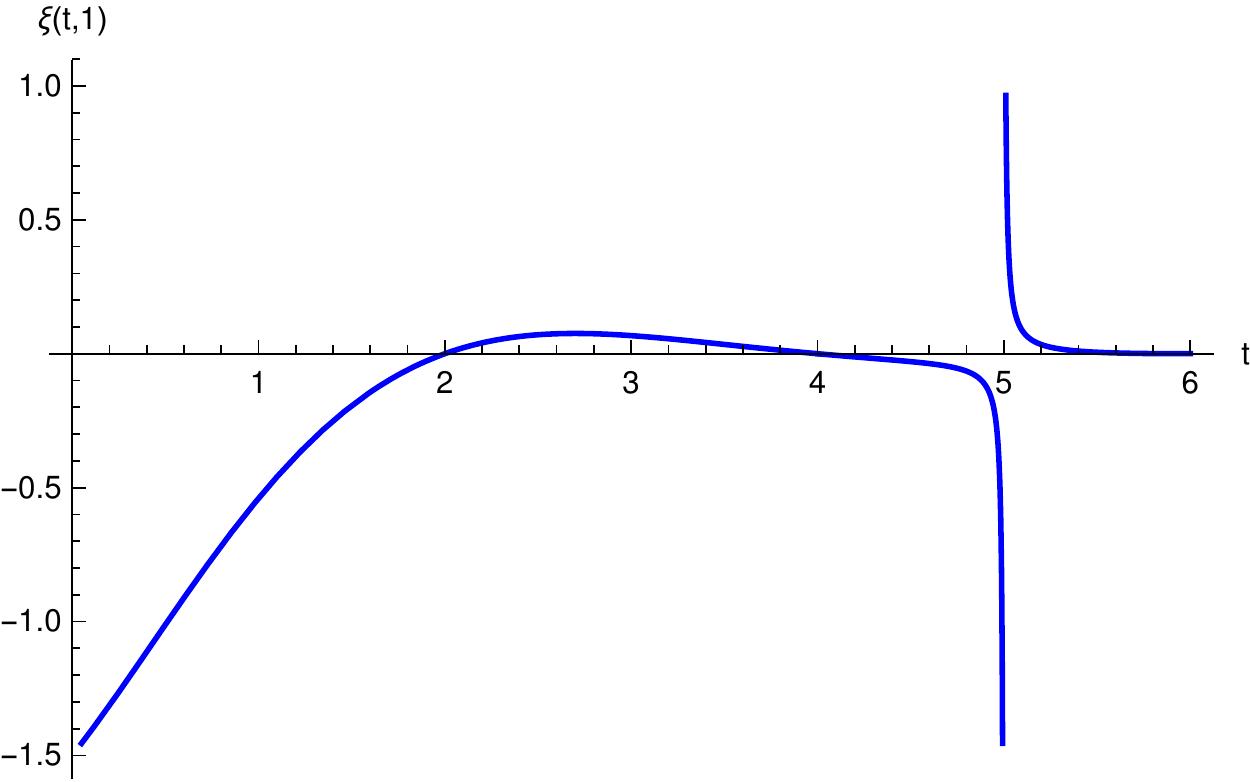}
\caption{The function $\xi(t,1)$.}
\label{fig:I1}
\end{figure}

Using the results of the previous section together with Eq.~\eqref{eq:ZK}, we can finally proceed to 
evaluating the $\beta$-function. First, we find that
\begin{equation}
G_1(K)= K + \frac{1}{N_f} \int_0^K 
\left( \xi_0(t) - \xi_0(0) + \xi(t,1) K 
+ \frac{\sigma_0(t)}{2} + v_0(t) \right) \mathrm{d}t.
\end{equation} 
Now, it is straightforward to compute the $\beta$-function:
\begin{equation}\label{eq:beta0}
\begin{split}
\beta(K)  =& K^2 + \frac{K^2}{N_f}
\left\{ - \xi_0(0) + \xi(K,1) + \frac{\sigma_0(K)}{2}
+ v_0(K)+ \int_0^K \xi(t,1) \mathrm{d}t 
\right\}.
\end{split}
\end{equation}
Recalling Eq.~\eqref{eq:xieps1} and using $\xi_0(0) = -\frac{3}{2}$, 
Eq.~\eqref{eq:beta0} can be further simplified to
\begin{equation}\label{eq:beta}
\frac{\beta(K)}{K^2} = 
1 + \frac{1}{N_f} \left\{ \frac{3}{2} + 
\int_0^K \xi(t,1) \mathrm{d}t \right\}.
\end{equation}
Finally, by comparison with Eq.~\eqref{eq:F0F1}, we see that $F_0 = 1$ and 
\begin{equation}
 F_1(K) = \frac{3}{2} + \int_0^K  \xi(t,1)  \mathrm{d}t.
\end{equation}
We plot the integrand, $\xi(t,1)$, in Fig.~\ref{fig:I1}. 
We have checked that our $\beta$-function agrees at the leading order in $N_f$ up to four-loop
level by comparing with the result of Ref.~\cite{Zerf:2017zqi}, and with the result extracted
from the critical exponents in Gross--Neveu--Yukawa model computed using a different 
technique~\cite{Gracey:2017fzu}.

Finally, let us comment on the pole structure: 
the integrand, $\xi(t,1)$, has the first pole occuring at $t=5$,
which results in a logarithmic singularity for $F_1(K)$ around $K=5$.  
Due to the sign of $\xi(t,1)$, we see that $F_1(K)$
approaches large negative values for $K\rightarrow5^-$. 
This suggests the existence of
a UV fixed point at $K_\text{UV} \lesssim 5$ such that $F_1(K_\text{UV}) = - N_f$.

\section{Conclusions}
\label{sec:concl}
We have computed the leading $1/N_f$ contribution 
for the $\beta$-function in Yukawa theory with $N_f$ fermion flavours
coupling to a real scalar. 
We obtained a closed form 
expression for the $\beta$-function up to order $\mathcal{O}(1/N_f)$.
This expression has a finite radius of convergence, and the first singularity occurs at $K=5$.

The present result adds an interesting ingredient to models with a large number of 
fermions, and makes a contribution to better understand the UV behaviour of gauge-Yukawa theories.


\section*{Acknowledgments}
We 
are grateful to John Gracey for 
bringing our attention the connection to the Gross--Neveu model.
We thank Florian Goertz and Valentin Tenorth for discussions and valuable comments. 

\appendix

\section{Loop integrals}
\label{sec:loops}
We here provide some explicit formulas. We follow closely the notations of Ref.~\cite{Grozin:2003ak}.
\subsection{The vertex and the fermion self-energy}
\label{sec:app1}
As shown in Eqs~\eqref{eq:Vn} and \eqref{eq:Sigmaexp}, the
1PI vertex and the fermion self-energy 
involve only the function $G(n_1,n_2)$, independently of
the number of bubbles. This
corresponds to the one-loop integral
\begin{equation}\label{eq:oneloop}
 \int \frac{ d^d k}{(2 \pi)^d} \frac{1}{D_1^{n_1} D_2^{n_2}} = 
 \mathrm{i}\frac{1}{(4 \pi)^{d/2}}
 (-p^2)^{d/2 - n_1 - n_2}(-1)^{n_1 + n_2}G(n_1,n_2),
\end{equation}
where $D_1 = (k+p)^2$ and $D_2 = k^2$. Explicitly, 
\begin{equation}
    \label{eq:G2}
 G(n_1,n_2) = 
 \frac{ \Gamma(-d/2 + n_1 + n_2)
 \Gamma(d/2 - n_1) \Gamma(d/2- n_2) }{\Gamma(n_1) \Gamma(n_2)
 \Gamma(d-n_1-n_2)}.
\end{equation}

\subsection{The scalar self-energy}
\label{sec:app2}
Unlike the 1PI vertex and the fermion self-energy, 
the $n$-loop contribution to the 
scalar self-energy, $\Pi_0$, indicated by
$\Pi^{(n)}$, cannot be written in terms of
$G(n_1,n_2)$ functions only.
In fact, $\Pi^{(n)}$ is given by ($n \geq 2$): 
\begin{equation}\label{eq:Piapp}
\begin{split}
p^2 \Pi^{(n)}(p^2,\epsilon)= & - (4 \pi^2)^2 (-1)^n
\left(\frac{1}{ (4 \pi)^{d/2-2} } \frac{G(1,1)}{2} \right)^{n-2} (-1)^\alpha
\int \frac{d^d k_1}{(2 \pi)^d} \int \frac{d^d k_2}{(2 \pi)^d} \\
& \bigg\{ \frac{6}{(p+k_1)^2 k_2^2((k_1-k_2)^2)^{1-\alpha}}
-\frac{2}{k_1^2(p+k_1)^2k_2^2((k_1-k_2)^2)^{-\alpha}} \\
& 
-\frac{2 p^2}{k_1^2 (p+k_1)^2 k_2^2 ((k_1-k_2)^2)^{1-\alpha}}
+\frac{2 p^2}{k_1^4 (p+k_1)^2 k_2^2((k_1-k_2)^2)^{-\alpha}} \\ 
& -
\frac{2 p^2}{k_1^2(k_1+p)^2(k_2+p)^2 k_2^2 ((k_1-k_2)^2)^{-\alpha}} \bigg\},
\end{split}
\end{equation}
where $\alpha = (n-2)(d/2-2) = -(n-2) \epsilon/2$.
Eq.~\eqref{eq:Piapp} requires two-loop integrals which
can be performed according to the
formula in Ref.~\cite{Grozin:2003ak}:
\begin{equation}
\int \frac{d^d k_1}{(2 \pi)^d} \int \frac{d^d k_2}{(2 \pi)^d}
\frac{1}{D_1^{n_1}D_2^{n_2}D_3^{n_3}D_4^{n_4}D_5^{n_5}}  =
(-1)^{1+\sum n_i} \frac{\pi^d(-p^2)^{d-\sum n_i}}{(2 \pi)^{2d}}  G(n_1,n_2,n_3,n_4,n_5), 
\end{equation}
where $D_1 = (k_1 + p)^2$, $D_2 = (k_2 + p)^2$, $D_3 = k_1^2$,
$D_4 = k_2^2$, $D_5 = (k_1-k_2)^2$. The functions $G(n_1,n_2,n_3,n_4,n_5)$
are symmetric with respect to the following index exchanges:
$ (1 \leftrightarrow 2, 3 \leftrightarrow 4)$
and $ (1 \leftrightarrow 3, 2 \leftrightarrow 4)$.
Moreover, they reduce 
to a product of $G(n_1,n_2)$ if at least one of the entries is zero:
\begin{equation}\label{eq:g1}
 G(n_1,n_2,n_3,n_4,0) = G(n_1,n_3)G(n_2,n_4),
\end{equation}
\begin{equation}\label{eq:g2}
 G(0,n_2,n_3,n_4,n_5) = G(n_3,n_5)G(n_2,n_3 + n_4 + n_5 -d/2).
\end{equation}
It turns out that the first four integrals
in Eq. \eqref{eq:Piapp} can always be written in terms of
$G(n_1,n_2)$ making use of Eqs~\eqref{eq:g1} and 
\eqref{eq:g2}.

However, the last integral in Eq.~\eqref{eq:Piapp} involves
$G(1,1,1,1,(n-2) \epsilon/2)$ and, for $n>2$, 
its expression can be obtained
in terms of hypergeometric functions
$\vphantom{F}_3F_2$
by means of the Gegenbauer technique~\cite{Kotikov:1995cw}.
We have evaluated the function $G(1,1,1,1,(n-2) \epsilon/2)$
recursively according to Eqs~(2.19) and (2.21) in Ref.~\cite{Grozin:2003ak}.

\bibliography{refs.bib}
\bibliographystyle{JHEP}

\end{document}